\documentclass[aps,pra,showpacs,amsmath,twocolumn,superscriptaddress]{revtex4-1}
\usepackage{amsmath,amssymb,graphicx}
\usepackage[final]{hyperref}
\usepackage{booktabs}

\hypersetup{
	colorlinks=true,       
	linkcolor=blue,          
	citecolor=blue,        
	filecolor=magenta,      
	urlcolor=blue
	}

\begin{document}

\title{Dispersion enhanced tunability of laser-frequency response to its
cavity-length change}

\author{Savannah L. Cuozzo}
\author{Eugeniy E. Mikhailov}
\email[eemikh@wm.edu]{}
\affiliation{Department of Physics, College of William $\&$ Mary, Williamsburg, Virginia 23187, USA}

\date{\today}

\begin{abstract}
	We report on the controllable response of the lasing 
	frequency to the cavity round-trip path change.
	This is achieved by modifying the dispersion  of the intracavity medium in
	the four-wave mixing regime in Rb.
	We can either increase the response by at least a factor of $2.7$ or
	drastically reduce it.
	The former regime is useful for sensitive
	measurements tracking the cavity round trip length and the latter
	regime is useful for precision metrology.
\end{abstract}


\maketitle

We control the response of the lasing
frequency to the laser cavity
length change on demand --- allowing for either dramatic enhancement
or 
suppression. 
The resonant frequency link to the cavity round trip path is
the foundation for optical precision measurements such as displacement tracking,
temperature sensing, optical rotation tracking~\cite{Chow1985RMP},
gravitational wave sensing~\cite{gwFirstDetection2016prl},
and refractive index change
sensing~\cite{laserFeedbackApplicationReview2017OE}. 
In other applications, the laser provides a stable frequency reference,
such as precision interferometry~\cite{advanceLIGO2018prd},
optical atomic clocks~\cite{opticalAtomicClocksReview2015}, and distance
ranging~\cite{distanceMeasurementReview2017}, where
the response of the lasing frequency to the cavity path length change should be
reduced. 
Our findings allow for improved laser assisted precision metrology and
potentially make lasers less bulky and immune to the 
environmental changes in real world applications.

\begin{figure}[h]
	\includegraphics[width=1.0\columnwidth]{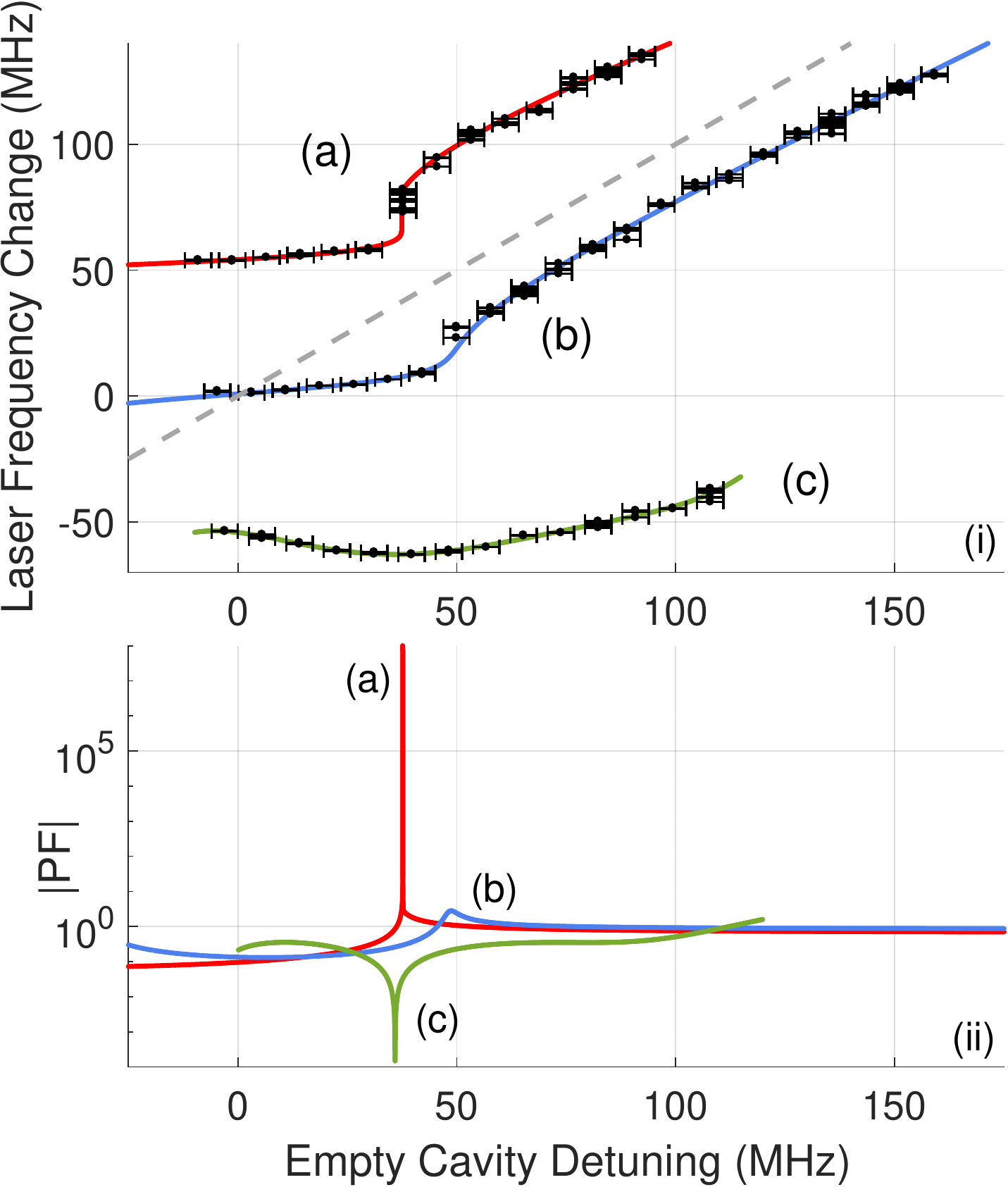}
	\caption{
		\label{fig:pulling_map_summary}
		(Color online)
		(i) is the experimental lasing frequency dependence on empty cavity detuning
		(round trip path change) in
		(a) bifurcating regime with estimated ultra high PF$>10^8$, 
		(b) high pulling regime with PF$=2.7\pm0.4$, 
		(c) enhanced stability regime where $|\mathrm{PF}| < 0.2$
		crossing 0.
		The solid lines (a and b) show our best fits of the laser frequency
		dependence using the model described by
		Eq.~\ref{eq:refraction_index_model};
		the (c) line is  the polynomial fit of the 5th degree.
		The straight dashed line shows the PF$=1$ dependence
		(i.e. for an empty cavity).  
		(ii) is the PF calculated based on the fits presented in (i).
	}
\end{figure}

The addition of a dispersive medium to a cavity modifies its frequency
response~\cite{shahriar2007pra_fast_gyro} to the
geometrical path change ($d p$) according to
\begin{equation}
	\label{eq:dispersive_cavity}
	d f_d  =
	- \frac{n}{n_\mathrm{g}}
	\frac{d p}{p_\mathrm{tot}} 
	f_0 ~,
\end{equation}
where $f_0$ is the original resonant frequency, $p_\mathrm{tot} = p_e +
p_d n$ is the total optical round-trip path of the cavity,
$p_d$ is the length of the dispersive element, $p_e$ is the length of the empty
(non-dispersive) part of the cavity,
$n$ is the refractive index,
and $n_\mathrm{g}$ is the generalized refractive group index given by
\begin{equation}
	\label{eq:group_index_cavity_modified}
	n_\mathrm{g} = n + \frac{ n p_d}{p_\mathrm{tot}} f_0 \frac{ \partial n}{\partial f}.
\end{equation}
We define
the pulling factor (PF) as the ratio of
dispersive to empty  (non-dispersive, $n_\mathrm{g} = n$) cavity
response for the same path change
\begin{equation}
	\label{eq:PF}
	\mathrm{PF} 
	\equiv
	\frac{d f_\mathrm{d}} { d f_\mathrm{e}} 
	= 
	\frac{n}{n_\mathrm{g}}.
\end{equation}
The PF is the figure of merit for the enhancement of the cavity response
relative to {\it canonical} lasers or passive cavities operating in 
the weak dispersion regime with $n_\mathrm{g} = n$.

We tune the PF in the range from -0.3 to at least $2.7\pm 0.4$ (see
Fig.~\ref{fig:pulling_map_summary}),
by tailoring the refractive index of our lasing medium.
This is the first demonstration of high and tunable PF in
the laser. We can also push our system response further and reach the
bifurcating regime.

Due to Kramers-Kronig relationship the
negative dispersion is accompanied by local absorption, so it is not
surprising  that so far the $PF > 1$ regime was experimentally
demonstrated only in passive, non-lasing
cavities~\cite{shahriar2007prl_wlc_demo, SmithPRA2016, smithOE2018_superluminal_cavity}
with PF$=363$.
For active cavities,
Yablon {\it et al.}~\cite{shahriar2015OE_towards_superl_laser_with_dip} {\em inferred} a
PF$\sim190$ via analysis of the lasing linewidth.
The increased stability regime (PF$<1$) was demonstrated in
lasing~\cite{Kutzke17OLtailorable_dispersion}
cavities with the smallest being PF$=1/663$~\cite{shahriar17OE_sensitivity_suppression}.
Superradiant (``bad-cavity'') lasers, where an atomic gain line is much narrower than a
cavity linewidth, exhibit ultra low
PF$<10^{-6}$~\cite{bohnet2012nature_stable_superradiant_laser,
norcia2018prx_stable_superradiant_laser}.
Our empty cavity linewidth is about 13~MHz which is larger than any atomic
decoherence time, so we operate in the ``bad-cavity'' regime. However,
unlike previously reported work in~\cite{bohnet2012nature_stable_superradiant_laser,
norcia2018prx_stable_superradiant_laser}, we can also achieve higher than
one PF.

\label{sec:theory}

\begin{figure}
	\includegraphics[width=0.99\columnwidth]{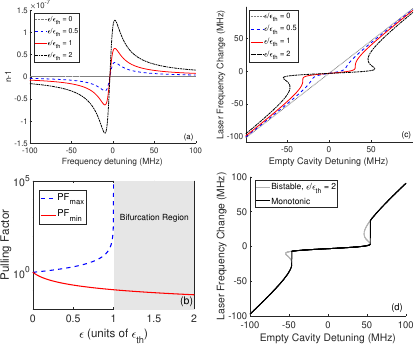}
	\caption{
		\label{fig:dispersion_analysis}
		(Color online)
		(a) Refractive index change ($n-1$);
		(b) dependence of maximum and minimal achievable PF on
		resonance strength;
		(c) laser frequency change 
		and (d) bifurcating behavior  
		as functions of detuning (or cavity path change).
		For all figures $\gamma$ is set to 6~MHz.
	}
\end{figure}

Similar to~\cite{shahriar16OCsubluminal_laser_modeling},
we present a simple model of the transmission or
amplification spectral line where the index of refraction has the 
dependence:
\begin{equation}
	\label{eq:refraction_index_model}
	n(f) = 1 + \epsilon 
	\frac{\gamma \Delta f } 
	{ \Delta f^2 + \gamma^2},
\end{equation}
where $\epsilon$ is the resonance strength, $\Delta f$ is the detuning from
the medium resonance frequency ($f_\mathrm{m}$), and $\gamma$ is the resonance
width, since $n(f)-1
\ll 10^{-5}$ for a vapor filled cavity.
For transmission or gain resonances with
$\epsilon > 0$, the minimum and maximum PF are
\begin{align}
	\label{eq:theoretical_pulling}
	\mathrm{PF}_{\mathrm{max}} &= \frac{1}{1-{\epsilon}/{
		\epsilon_{\mathrm{th}}}}
			~~~\mathrm{at}~~~ \Delta f = \pm  \sqrt{3}\gamma,\\
	\mathrm{PF}_{\mathrm{min}} &= \frac{1}{1+{8\epsilon}/{ \epsilon_{\mathrm{th}}}}
	~~\mathrm{at}~~~ \Delta f = 0
\end{align}
where 
\begin{align}
	\label{eq:epsilon_threshold}
	~ \epsilon_{\mathrm{th}} &= \frac{8\gamma}{f_\mathrm{m}}
	\frac{p_\mathrm{tot}}{p_\mathrm{d}}
\end{align}
is the bifurcating threshold resonance strength.

The analysis of the dispersion  (Eq.~\ref{eq:refraction_index_model})
and its influence on the resonant frequency
of the cavity and PF is shown in Fig.~\ref{fig:dispersion_analysis}.
As expected, the amplification line has positive dispersion
on resonance (see Fig.~\ref{fig:dispersion_analysis}a). Positive dispersion is associated with a large and positive group index, which results in weak
dependence (low pulling factor) of the lasing frequency on the cavity path change (empty cavity
detuning), as shown in Fig.~\ref{fig:dispersion_analysis}b.
Away from resonance, the dispersion is negative leading to 
high PFs, as shown in Fig.~\ref{fig:dispersion_analysis}b.
The stronger the
amplification ($\epsilon$) the smaller the PF$_\mathrm{min}$ is at the
center of the resonance,
as shown in Fig.~\ref{fig:dispersion_analysis}b. 
Consequently, the PF$_\mathrm{max}$ continuously grows and
reaches infinity at $\epsilon=\epsilon_\mathrm{th}$ where the resonant frequency
bifurcates (see Fig.~\ref{fig:dispersion_analysis}b).

To track dependence of the cavity resonant frequency on the cavity path
length, we solve:
\begin{equation}
	p_{\mathrm{tot}} = m \frac{c}{f_d},
\end{equation}
where 
$m$ is the fixed mode number and
$c$ is the speed of light in vacuum. 
In experiment, it is easier to track the empty cavity detuning
(i.e. resonance frequency change, $\Delta f_\mathrm{e}$),
which is directly linked to the cavity path
change via Eq.~\ref{eq:dispersive_cavity} with $n_\mathrm{g}=n$.
The resulting dependencies are shown in
Fig.~\ref{fig:dispersion_analysis}c.

If the negative
dispersion is strong enough, the group index could be negative.
This
would lead to negative PF and to negative dependence of the
lasing frequency on the cavity detuning (see line corresponding
$\epsilon=2  \epsilon_{\mathrm{th}}$ in Fig.~\ref{fig:dispersion_analysis}c).
This behavior is
nonphysical, since it corresponds to a
bifurcation~\cite{shahriar16OCsubluminal_laser_modeling}:
multiple lasing frequencies for the same cavity detuning. Consequently, the laser would 
`jump' to avoid negative  PF  region and preserve the monotonic behavior, as shown in
Fig.~\ref{fig:dispersion_analysis}d
and experimentally in
Fig.~\ref{fig:pulling_map_summary}(i)a.

The most important conclusion from the amplifying line analysis is that high
pulling (response enhancement) regions exist
slightly away from the {\em gain} resonance. The precursor of such regime is a reduced PF region in close vicinity to the resonance.
The off resonance behavior was overlooked in the literature,
while it  actually provides the road to high PF.
Away from the amplification resonance, the system
still has enough gain to sustain lasing, and yet it still has large
negative dispersion
(see Fig.~\ref{fig:dispersion_analysis}a). As detuning from the resonance
increases, the dispersion becomes 
negligible, PF approaches unity (see Fig.~\ref{fig:dispersion_analysis}c
and experimental data in
Fig.~\ref{fig:pulling_map_summary} a and b),
amplification drops and eventually lasing ceases.


\begin{figure}
	\includegraphics[width=0.7\columnwidth]{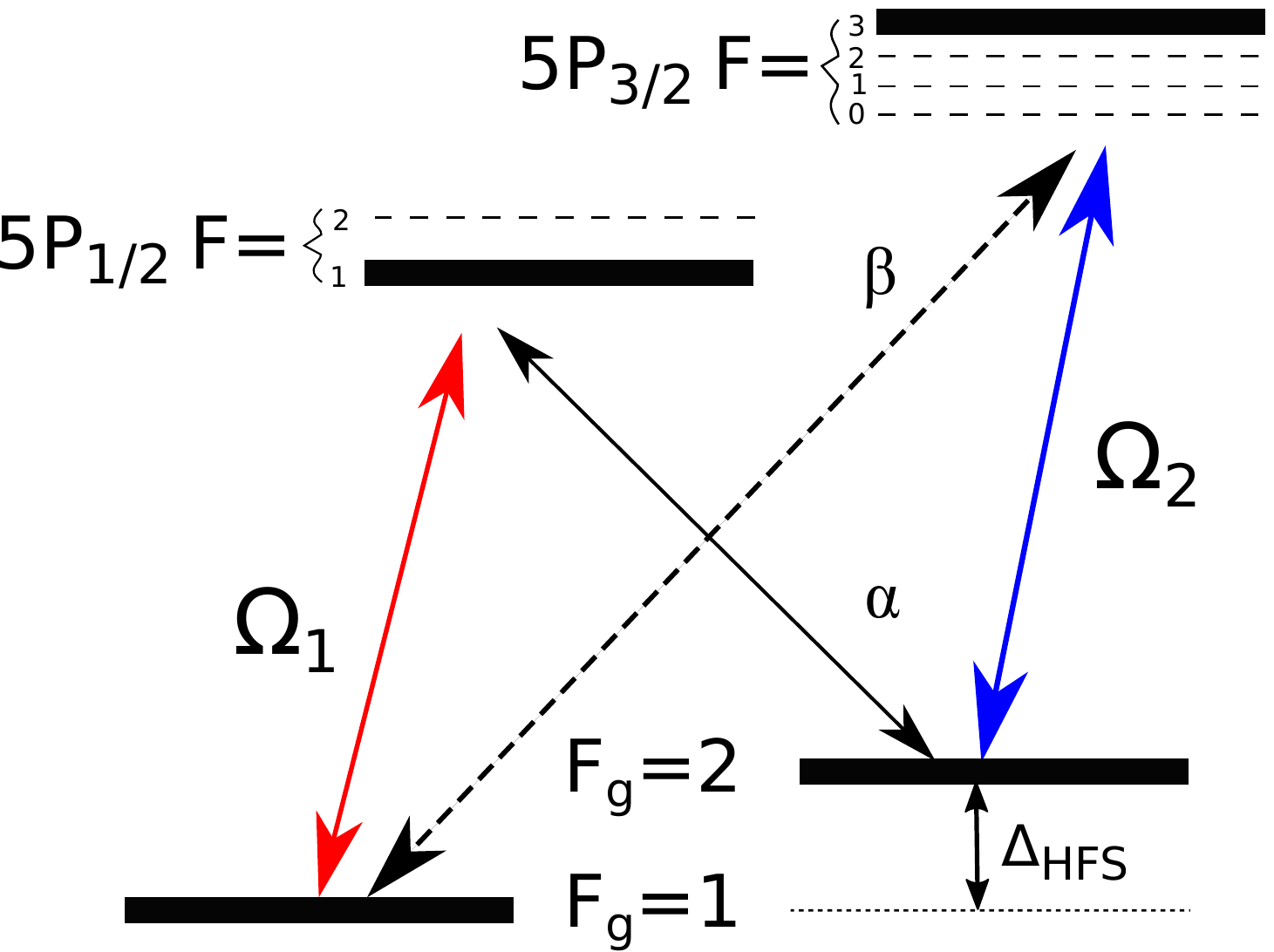}
	\caption{
		\label{fig:N_levels_diagram}
		(Color online)
		Schematic diagram of interacting light fields and relevant
		$^{87}$Rb levels.
	}
\end{figure}

\begin{figure}
	\includegraphics[width=0.9\columnwidth]{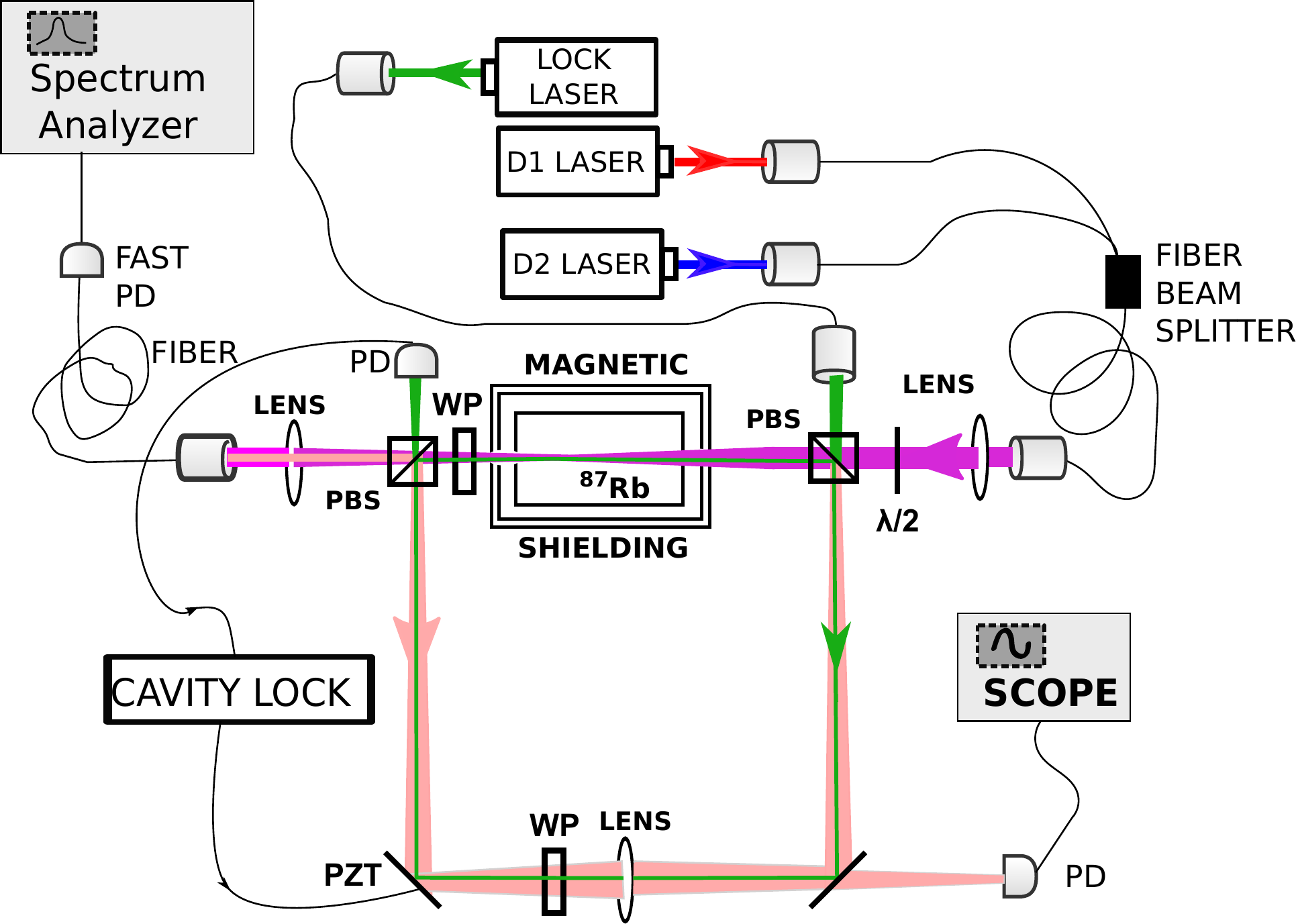}
	\caption{
		\label{fig:setup}
		(Color online)
		Schematic diagram of the setup.
		Labels are: PD is photo detector,
		WP is wave plate, 
		$\lambda/2$ is half wavelength wave plate,
		PBS is polarizing beam splitter,
		and PZT is piezoelectric transducer.
	}
\end{figure}

To experimentally demonstrate the modified  lasing response  to the cavity
path change,
one needs a narrow gain line to achieve the highest positive dispersion. We utilized the 
N-level pumping scheme depicted in Fig.~\ref{fig:N_levels_diagram}. The
theory and preliminary experimental study of this arrangement are covered
in
references~\cite{Kutzke17OLtailorable_dispersion,
mikhailovOE2014fwm_in_ring_cavity, mikhailov2013jmo_fast_N_scheme}.
The strong pumping
field $\Omega_1$ creates a transmission line for the field $\alpha$ due to
electromagnetically induced transparency. However, the $\Omega_1$ field
alone is not enough to create the amplification. To create the gain for the
$\alpha$ field, we apply another strong repumping field ($\Omega_2$).
There is also gain for the $\beta$ field, which completes
the four-wave mixing arrangement of fields $\Omega_1$, $\Omega_2$, $\alpha$,
and $\beta$. But the cavity is tuned to sustain lasing only for $\alpha$.

Our lasing cavity is similar to the one used
in~\cite{Kutzke17OLtailorable_dispersion}.
The ring cavity is made of two polarizing beam splitters (PBS) and two flat
mirrors. The round trip path of the cavity is 80~cm.
A 22~mm long Pyrex cylindrical cell with anti-reflection coatings on its windows
is placed between the two PBSs and filled with isotopically pure $^{87}$Rb.
The cell is encased in a 3-layer magnetic 
shield and its temperature is set to 100 degrees C.
The optical stability of the cavity is increased by adding a 
30~cm focal length lens placed between the two mirrors. This lens also  places the cavity's
mode waist inside the ${}^{87}$Rb cell. 

To produce experimental data sets (a) and (b) shown in the
Fig.~\ref{fig:pulling_map_summary},
two pump lasers are tuned near
D1 (795~nm) and D2 (780~nm) corresponding to $\Omega_1$ and $\Omega_2$ fields
in Fig.~\ref{fig:N_levels_diagram}.
The pump fields are coupled to a fiber beam splitter and amplified by
a  solid state tapered amplifier 
to powers ranging between 100~mW for set (a) and 170~mW for set (b),
and then injected into a ring cavity through a polarizing beam splitter
(PBS).
The D1 laser is tuned 700~MHz below
the $5S_{1/2}$F${}_\mathrm{g}$=1 $\rightarrow$ $5P_{1/2}$F=1 transition,
and D2 is set to 500~MHz below
the $5S_{1/2}$F${}_\mathrm{g}=2$ $\rightarrow$  $5P_{3/2}$F=3  $^{87}$Rb transition,
as seen in Fig.~\ref{fig:N_levels_diagram}. 
They provide amplification for
fields $\alpha$ and $\beta$, which  are generated orthogonal to pump
fields polarization.
Only the $\alpha$ field resonantly circulates in the cavity, since the pumps
exit the cavity via the second PBS and $\beta$ is kept off-resonance with the
cavity.

Since the D1 pump laser is fixed,
the beatnote of the pump ($\Omega_1$) and the lasing field ($\alpha$)
with its frequency close to $^{87}$Rb hyperfine splitting
($\Delta_{\mathrm{HFS}} \approx  6.8$~GHz) is related to the frequency
of the ring cavity laser and allows us to
monitor the dispersive laser frequency change ($\Delta f_d$).
We control the cavity length by locking it to an auxiliary laser (called
the lock laser) with a wavelength of 795~nm that is far detuned from 
any atomic resonances and senses a ``would be empty'' (dispersion-free)
cavity detuning ($\Delta f_e$).
This lock laser beam counter propagates relative the pump beams and the
lasing field to avoid contaminating the detectors monitoring the ring
cavity lasing.
Two wave plates (WP) are placed inside the cavity. One is to spoil
polarization of the lock field and allow it to circulate in the cavity. The
other rotates the lasing field polarization by a small amount. This
allows it to exit the cavity and mix in with the pump field on the
fast photodetector.

\label{sec:results}

The maximum response has
the lower bound of
PF${}_\mathrm{max}=1.1 \times 10^8$ at the 90\% confidence level
for the data set (a), shown
in Fig.~\ref{fig:pulling_map_summary}.
The upper bound for PF${}_\mathrm{max}$ is infinity since the data set belongs to the
bifurcating regime. However, one can smoothly approach this limit by
carefully controlling the cavity detuning as our analysis shows in
Fig.~\ref{fig:pulling_map_summary}(ii)a.
The PF${}_\mathrm{min}$ range is $(0.08$ to $0.10)$ for this data set. 

We can avoid bifurcation by increasing the pumps' powers (i.e., we increase
$\gamma$ via power broadening), as shown in the data
set (b) of Fig.~\ref{fig:pulling_map_summary}. This data demonstrates
PF$_\mathrm{max}$ in the range (2.3 to 3.2). Also, the range of
detuning with PF$>1$ is wider.
To estimate confidence bounds, we use the
modified smoothed bootstrap method~\cite{bootstrapBookbyEfron}.

We are able
to make our dispersive laser insensitive to its path change,
as shown in data set (c) of Fig.~\ref{fig:pulling_map_summary}.
We tune tune 
the D1 laser to 400~MHz above 
the $5S_{1/2}$F${}_\mathrm{g}$=1 $\rightarrow$ $5P_{1/2}$F=1 transition,
and keep  D2  at  500~MHz below
the $5S_{1/2}$F${}_\mathrm{g}=2$ $\rightarrow$  $5P_{3/2}$F=3  $^{87}$Rb
transition, while maintaining combined pump power at 95~mW.
Assuming a smooth dependence on the empty cavity detuning, the PF at the
bottom of the U-like curve is exactly zero, as the laser frequency
decreases and then increases, 
while the cavity path (the auxiliary laser detuning) changes monotonically.
Our model governed by Eq.~\ref{eq:refraction_index_model}
cannot explain the arching behavior,
since it does not account for the
dependence of the dispersion on the lasing power. However, a more complete
model which solves density matrix equations of the N-level scheme predicted
such a possibility~\cite{Kutzke17OLtailorable_dispersion}.

There is an ongoing debate whether or not the modified cavity response leads to
improved sensitivity (signal to noise ratio) of path-change sensitive
detectors. However,
laser-based sensors in certain applications
might benefit either from enhanced PF$ > 1$ (for example
gyroscopes~\cite{shahriar2007pra_fast_gyro}) or reduced PF$< 1$, since
sensitivity, i.e. the ratio of the response to the lasing linewidth
(uncertainty),
scales as
1/PF~\cite{kuppensPRL1994linewidthInBadCavity,bohnet2012nature_stable_superradiant_laser,henry1982ieee}.
The tunability and
versatility of our system allows to probe either case.


In conclusion,
we achieved
about $2.7 \pm 0.4$ increase of the laser response to the cavity-path length
change
relative to canonical lasers.
We also can significantly
reduce the response, making our laser vibration insensitive.
These findings broadly impact the fields of
laser sensing and metrology, including laser ranging, laser gyroscopes,
vibrometers, and laser frequency standards.

EEM  and SLC are thankful to the support of
Virginia Space Grant Consortium provided by grants 17-225-100527-010 and NNX15AI20H.
We would like to thank M. Simons, O. Wolfe, and D. Kutzke for assembling
the earlier prototype of our laser.
We also thank S. Rochester and D. Budker for their earlier work on N-level
theoretical model.

%


\end{document}